\documentstyle[twoside,psfig]{article}
\newcommand{\be}{\begin{equation}}
\newcommand{\ee}{\end{equation}}
\newcommand{\beq}{\begin{equation}}
\newcommand{\eeq}{\end{equation}}
\newcommand{\p}{\partial}
\newcommand{\bea}{\begin{eqnarray}}
\newcommand{\eea}{\end{eqnarray}}
\catcode`\@=11
\long\def\@makefntext#1{
\protect\noindent \hbox to 3.2pt {\hskip-.9pt  
$^{{\eightrm\@thefnmark}}$\hfil}#1\hfill}		

\def\thefootnote{\fnsymbol{footnote}}
\def\@makefnmark{\hbox to 0pt{$^{\@thefnmark}$\hss}}	
	
\def\ps@myheadings{\let\@mkboth\@gobbletwo
\def\@oddhead{\hbox{}
\rightmark\hfil\eightrm\thepage}   
\def\@oddfoot{}\def\@evenhead{\eightrm\thepage\hfil
\leftmark\hbox{}}\def\@evenfoot{}
\def\sectionmark##1{}\def\subsectionmark##1{}}

\oddsidemargin=\evensidemargin
\renewcommand{\thefootnote}{\fnsymbol{footnote}}

\newcounter{sectionc}\newcounter{subsectionc}\newcounter{subsubsectionc}
\renewcommand{\section}[1] {\vspace{12pt}\addtocounter{sectionc}{1} 
\setcounter{subsectionc}{0}\setcounter{subsubsectionc}{0}\noindent 
	{\tenbf\thesectionc. #1}\par\vspace{5pt}}
\renewcommand{\subsection}[1] {\vspace{12pt}\addtocounter{subsectionc}{1} 
	\setcounter{subsubsectionc}{0}\noindent 
	{\bf\thesectionc.\thesubsectionc. {\kern1pt \bfit #1}}\par\vspace{5pt}}
\renewcommand{\subsubsection}[1] {\vspace{12pt}\addtocounter{subsubsectionc}{1}
	\noindent{\tenrm\thesectionc.\thesubsectionc.\thesubsubsectionc.
	{\kern1pt \tenit #1}}\par\vspace{5pt}}
\newcommand{\nonumsection}[1] {\vspace{12pt}\noindent{\tenbf #1}
	\par\vspace{5pt}}

\newcounter{appendixc}
\newcounter{subappendixc}[appendixc]
\newcounter{subsubappendixc}[subappendixc]
\renewcommand{\thesubappendixc}{\Alph{appendixc}.\arabic{subappendixc}}
\renewcommand{\thesubsubappendixc}
	{\Alph{appendixc}.\arabic{subappendixc}.\arabic{subsubappendixc}}

\renewcommand{\appendix}[1] {\vspace{12pt}
        \refstepcounter{appendixc}
        \setcounter{figure}{0}
        \setcounter{table}{0}
        \setcounter{lemma}{0}
        \setcounter{theorem}{0}
        \setcounter{corollary}{0}
        \setcounter{definition}{0}
        \setcounter{equation}{0}
        \renewcommand{\thefigure}{\Alph{appendixc}.\arabic{figure}}
        \renewcommand{\thetable}{\Alph{appendixc}.\arabic{table}}
        \renewcommand{\theappendixc}{\Alph{appendixc}}
        \renewcommand{\thelemma}{\Alph{appendixc}.\arabic{lemma}}
        \renewcommand{\thetheorem}{\Alph{appendixc}.\arabic{theorem}}
        \renewcommand{\thedefinition}{\Alph{appendixc}.\arabic{definition}}
        \renewcommand{\thecorollary}{\Alph{appendixc}.\arabic{corollary}}
        \renewcommand{\theequation}{\Alph{appendixc}.\arabic{equation}}
        \noindent{\tenbf Appendix \theappendixc #1}\par\vspace{5pt}}
\newcommand{\subappendix}[1] {\vspace{12pt}
        \refstepcounter{subappendixc}
        \noindent{\bf Appendix \thesubappendixc. {\kern1pt \bfit #1}}
	\par\vspace{5pt}}
\newcommand{\subsubappendix}[1] {\vspace{12pt}
        \refstepcounter{subsubappendixc}
        \noindent{\rm Appendix \thesubsubappendixc. {\kern1pt \tenit #1}}
	\par\vspace{5pt}}

\newcommand{\textlineskip}{\baselineskip=13pt}
\newcommand{\smalllineskip}{\baselineskip=10pt}

\def\eightcirc{
\begin{picture}(0,0)
\put(4.4,1.8){\circle{6.5}}
\end{picture}}
\def\eightcopyright{\eightcirc\kern2.7pt\hbox{\eightrm c}} 

\newcommand{\copyrightheading}[1]
	{\vspace*{-2.5cm}\smalllineskip{\flushleft
	{\footnotesize Modern Physics Letters A, #1}\\
	{\footnotesize $\eightcopyright$\, World Scientific Publishing
	 Company}\\
	 }}

\newcommand{\publisher}[2]{{\begin{center}\footnotesize\smalllineskip 
	Received #1\\
	Revised #2
	\end{center}
	}}

\def\abstracts#1#2#3{{
	\centering{\begin{minipage}{4.5in}\footnotesize\baselineskip=10pt
	\parindent=0pt #1\par 
	\parindent=15pt #2\par
	\parindent=15pt #3
	\end{minipage}}\par}} 

\def\keywords#1{{
	\centering{\begin{minipage}{4.5in}\footnotesize\baselineskip=10pt
	{\footnotesize\it Keywords}\/: #1
	 \end{minipage}}\par}}

\newcommand{\bibit}{\nineit}
\newcommand{\bibbf}{\ninebf}
\renewenvironment{thebibliography}[1]
	{\frenchspacing
	 \ninerm\baselineskip=11pt
	 \begin{list}{\arabic{enumi}.}
        {\usecounter{enumi}\setlength{\parsep}{0pt}     
	 \setlength{\leftmargin 12.7pt}{\rightmargin 0pt} 
         \setlength{\itemsep}{0pt} \settowidth
	{\labelwidth}{#1.}\sloppy}}{\end{list}}

\newcounter{itemlistc}
\newcounter{romanlistc}
\newcounter{alphlistc}
\newcounter{arabiclistc}

\newcommand{\fcaption}[1]{
        \refstepcounter{figure}
        \setbox\@tempboxa = \hbox{\footnotesize Fig.~\thefigure. #1}
        \ifdim \wd\@tempboxa > 5in
           {\begin{center}
        \parbox{5in}{\footnotesize\smalllineskip Fig.~\thefigure. #1}
            \end{center}}
        \else
             {\begin{center}
             {\footnotesize Fig.~\thefigure. #1}
              \end{center}}
        \fi}

\newcommand{\tcaption}[1]{
        \refstepcounter{table}
        \setbox\@tempboxa = \hbox{\footnotesize Table~\thetable. #1}
        \ifdim \wd\@tempboxa > 5in
           {\begin{center}
        \parbox{5in}{\footnotesize\smalllineskip Table~\thetable. #1}
            \end{center}}
        \else
             {\begin{center}
             {\footnotesize Table~\thetable. #1}
              \end{center}}
        \fi}

\def\@citex[#1]#2{\if@filesw\immediate\write\@auxout
	{\string\citation{#2}}\fi
\def\@citea{}\@cite{\@for\@citeb:=#2\do
	{\@citea\def\@citea{,}\@ifundefined
	{b@\@citeb}{{\bf ?}\@warning
	{Citation `\@citeb' on page \thepage \space undefined}}
	{\csname b@\@citeb\endcsname}}}{#1}}

\newif\if@cghi
\def\cite{\@cghitrue\@ifnextchar [{\@tempswatrue
	\@citex}{\@tempswafalse\@citex[]}}
\def\citelow{\@cghifalse\@ifnextchar [{\@tempswatrue
	\@citex}{\@tempswafalse\@citex[]}}
\def\@cite#1#2{{$\null^{#1}$\if@tempswa\typeout
	{IJCGA warning: optional citation argument 
	ignored: `#2'} \fi}}

\def\pmb#1{\setbox0=\hbox{#1}
	\kern-.025em\copy0\kern-\wd0
	\kern.05em\copy0\kern-\wd0
	\kern-.025em\raise.0433em\box0}

\def\fnt#1#2{\footnotetext{\kern-.3em
	{$^{\mbox{\scriptsize #1}}$}{#2}}}

\def\fpage#1{\begingroup
\voffset=.3in
\thispagestyle{empty}\begin{table}[b]\centerline{\footnotesize #1}
	\end{table}\endgroup}

\def\runninghead#1#2{\pagestyle{myheadings}
\markboth{{\protect\footnotesize\it{\quad #1}}\hfill}
{\hfill{\protect\footnotesize\it{#2\quad}}}}
\font\tenrm=cmr10
\font\tenit=cmti10 
\font\tenbf=cmbx10
\font\bfit=cmbxti10 at 10pt
\font\ninerm=cmr9
\font\nineit=cmti9
\font\ninebf=cmbx9
\font\eightrm=cmr8

\def\qed{\hbox{${\vcenter{\vbox{			
   \hrule height 0.4pt\hbox{\vrule width 0.4pt height 6pt
   \kern5pt\vrule width 0.4pt}\hrule height 0.4pt}}}$}}

\renewcommand{\thefootnote}{\fnsymbol{footnote}}	

\begin{document}
\setlength{\textheight}{7.7truein}  

\runninghead{Naked singularities and the Wilson loop $\ldots$}
{Naked singularities and the Wilson loop  $\ldots$}

\normalsize\textlineskip
\thispagestyle{empty}
\setcounter{page}{1}

\copyrightheading{}			

\vspace*{0.88truein}

\fpage{1}
\centerline{\bf NAKED SINGULARITIES AND THE WILSON LOOP}
\baselineskip=13pt
 
\vspace*{0.37truein}

\centerline{\footnotesize VIQAR HUSAIN }
\baselineskip=12pt
\centerline{\footnotesize\it Department of Mathematics and Statistics, University
of New Brunswick}
\baselineskip=10pt
\centerline{\footnotesize\it Fredericton, NB Canada E3B 5A3} 
 
\vspace*{0.225truein}

\publisher{(received date)}{(revised date)}

\vspace*{0.21truein}
\abstracts{We give observations about dualities where one of the dual theories 
is geometric. These are illustrated with a duality between the simple harmonic 
oscillator and a topological field theory. We then discuss the Wilson loop  
in the context of the AdS/CFT duality. We show that the Wilson loop calculation
for certain asymptotically AdS scalar field spacetimes with naked singularities 
gives results qualitatively similar to that for the AdS black hole. In particular, 
it is apparent that (dimensional) metric parameters in the singular spacetimes  
permit a ``thermal screening'' interpretation for the quark potential in the boundary 
theory, just like black hole mass. This suggests that the Wilson loop calculation 
merely captures metric parameter information rather than true horizon 
information.}{}{}

\vspace*{10pt}


\vspace*{1pt}\textlineskip	

\section{Introduction}	
\vspace*{-0.5pt}
\noindent
The basic idea underlying the term `duality'' may be stated as follows: 
Two distinct classical theories, prescribed by actions based on different 
sets of fields, not necessarily in the same spacetime dimension, may be 
equivalent at the quantum level. This means that it is possible to establish 
a correspondence between observables and states in the respective quantum 
theories. If the difference in spacetime dimensions of the dual theories 
is $\pm 1$, then the term ``holography'' is used to capture the general 
idea that all physical information of the higher dimensional theory is 
potentially manifested in one lower dimension. 

That such correspondences may be possible should not come as 
a surprise since quantum theories know about spacetime dimension only 
through the choice of representation chosen for classical observables. 
In particular the coordinate and momentum representations are the specific 
choices which inject spacetime dimension information into quantum theory. 
Their use is not necessary since even the simple harmonic oscillator may 
be quantized algebraically via the creation and annihilation operators. 

In theories with large gauge invariance groups, particularly general 
coordinate invariance, the interest lies in identifying and 
using gauge invariant observables for quantization. In such theories 
it is not appropriate to quantize based on the fundamental
Poisson bracket, but rather on a suitable Poisson algebra of gauge 
invariant observables.\cite{ish} For generally covariant theories without 
matter fields, such observables are necessarily non-local. With matter,  
local observables may be constructed by using combinations of matter 
and geometric variables such that the matter acts as a reference 
system to locate spacetime points.\cite{kuchar} In either case, the emphasis 
is on first finding gauge invariant classical observables, and attempting 
to represent these as operators in a quantum theory.\footnote{An alternative 
is to construct a quantum theory by first eliminating all gauge 
invariances at the classical level; the remaining problem here is 
whether different gauges lead to unitarily equivalent quantum theories.}

Dualities are potentially very useful if they allow the probing of 
one quantum theory using another which is better understood. 
In this sense dualities may be viewed as a means for obtaining 
new ``collective variables'' for studying a given theory in 
spacetime or energy domains where other variables reach their 
limits of usefulness. Of particular interest is the question 
of whether dualties can provide insights for quantum gravity. The 
AdS/CFT correspondence\cite{malda,review} in principle has the potential 
to do so. However to date it has yielded little insight into what happens 
to spacetime at the quantum level, or the role played by diffeomorphism 
invariance in the dual CFT. For this reason it may be useful to search 
for duals of geometric theories simpler than general relativity or 
supergravity. An interesting probe of this correspondence would be 
to seek manifestations in the CFT of interesting metric structures 
such as event horizons. A more general question in this regard is 
understanding of the role played in the CFT by metric parameters in 
asymptotically AdS spacetimes.

In this paper I probe some of these questions. I first describe a 
rather explicit duality between the simple harmonic oscillator and a 
topological field theory in four dimensions. Following this I discuss the 
AdS/CFT duality between gravity on 5-dimensional AdS spacetime and 
4-dimensional conformal Yang-Mills theory, in particular the Wilson loop 
calculation. This calculation may be done explicitly for asymptotically AdS 
scalar field spacetimes with naked singularities. The results are remarkably 
similar to those for the AdS black hole. I discuss the interpretation of this 
result with emphasis on whether the calculation captures true horizon and 
temperature information in the CFT.  
 
\setcounter{footnote}{0}
\renewcommand{\thefootnote}{\alph{footnote}}

\section{A Toy Duality}
\noindent
Consider the 4-dimensional topological field theory given by the action
\be 
S = \int_M B\wedge F(A). 
\ee 
The fields are the Abelian connection $A$ and the two-form 
$B$, with $F(A) = dA$. This action is invariant under diffeomorphisms, 
$U(1)$ gauge transformations, and $B \rightarrow B + d\Lambda$ for one 
forms $\Lambda$ (for vanishing surface terms). 
Consider the Hamiltonian quantization of this action for manifolds 
$M \sim \Sigma \times R$ where $\Sigma$ is compact without boundary.
\footnote{Quantization of models of this type have been discussed before.  
The Hamiltonian approach followed here is a review of earlier work by the 
author.\cite {vtop}} Since the action is first order, it is easy to put 
into canonical from:
\be 
S = \int_{\Sigma \times R} \epsilon^{0abc}\left[ B_{ab}\partial_0A_c + 
2B_{0a}F_{bc} - B_{ab}\partial_c A_0  \right].
\ee
Thus the canonical coordinates are $(A_a,E^a)$, where $E^a = \epsilon^{0abc} 
B_{bc}$. The Hamiltonian is a linear combination of the constraints 
\be
F_{ab} = 0, \ \ \ \ \ \ \ \ \ \partial_aE^a = 0.
\ee
These constraints are first class and generate the gauge transformations 
of the theory. Note that the Hamiltonian is a linear combination of 
the two contraints as expected for a generally covariant theory,  
and that spatial diffeomorphisms are generated by the combination
$A_a\partial_bE^b + E^bF_{ab}$.  

Since the constraints generate gauge transformations, gauge invariant 
observables ${\cal O}(E,A)$ are defined by the  Poisson 
bracket conditions 
\be 
\left\{ {\cal O}(E,A), C(E,A) \right\} = 0,
\ee
where $C$ denotes the two constraints. In the present case the basic
observables satisfying this condition are 
\be 
T^0(A,\gamma) = {\rm exp}\left[\int_\gamma ds \dot{\gamma}^a A_a \right],
 \ \ \ \ \ \ \ \ \ 
T^1(E,S) = \int_S d^2\sigma\ n_aE^a,
\ee
which are parametrized by loops $\gamma$ and surfaces $S$, $n_a$ is a 
one form field defining the surface $S$ ($\epsilon^{0abc}n_c$ is the area 
2-form and $\dot{\gamma}^a$ is tangent vector to the loop $\gamma$). These 
observables satisfy the Poisson algebra
\be 
\left\{ T^0(A,\gamma), T^0(A,\beta)\right\} = 0, \ \ \ \ \ \ \ \ \ \ 
\left\{ T^1(E,S), T^1(E,S') \right\} = 0,
\ee
\be
 \left\{T^0(A,\gamma), T^1(E,S) \right\} = c(\gamma, S) T^0(A,\gamma)
 \label{nonc}
\ee
where 
\be
c(\gamma,S) = \int ds\int d^2\sigma\ \dot{\gamma}^an_a\delta^3(\gamma(s) - S(\sigma)).
\ee 
The last Poisson bracket vanishes if the loop lies in the surface. 

On the constraint surface the observables $T^0$ and $T^1$ depend only on 
the non- contractible loops and surfaces in $\Sigma$. Thus they capture 
topological information about $\Sigma$. To proceed further we must fix the 
topology, which determines the number of independent observables, and hence 
degrees of freedom. For the case $\Sigma \sim S^1\times S^2$ there is exactly 
one non-contractible loop and surface, for which $c(\gamma,S)=1$. Thus there 
are two degrees of freedom, and the Poisson algebra (\ref{nonc}) becomes 
\be 
 \left\{ T^0, T^1\right\} = T^0.
\ee  
A quantum theory can be constructed by representing this algebra on a state space 
of ``occupation numbers'' by defining 
\be 
\hat{T}^0|n> = \sqrt{n}|n-1> \ \ \ \ \ \ \ \ \ \ \ \ \hat{T}^1|n> = n|n>.
\ee
The commutator algebra following from these defintions has the appropriate classical 
limit. This establishes the duality with the harmonic oscillator, with $T^1$ 
corresponding to the ``composite'' number operator $a^\dagger a$. 

This procedure may be applied in other dimensions and spacetime topologies. The number 
of degrees of freedom depend on the topology. There is an extension of the observables 
to the non-Abelian case, which is a bit more involved, and the algebra of observables 
does not correspond to the harmonic oscillator.\cite{vtop}

The basic lesson from this example is that dualities may turn out to be of no 
more than mathematical interest. The lesson for quantum gravity is that 
even if this type of Hamiltonian procedure can be carried out to completion 
starting from classical general relativity (without matter), the task of 
extracting a classical spacetime would be formidable. This is because 
fully gauge invariant observables, being non-local, would not carry any local 
spacetime information. This problem may be eased by solving the Hamiltonian
constraint at the classical level via gauge fixing to a preferred classical time
coordinate, or alternatively, by incorporating matter and using it to locate 
spacetime points in a diffemorphism invariant manner.\cite{kuchar}

\section{The Wilson loop in AdS/CFT}
\noindent
Dualities between theories with infinitely many degrees of freedom 
are clearly of much more interest than quantum mechanical models 
of the type discussed in the previous section. An early example is 
the Thirring model and its bosonic dual. 

A potential basis for constructing dual theories is the basic observation 
that quantum field theories on a fixed background spacetime carry 
representations of the spacetime isometry group. Since the possibility 
exists that spacetimes of different dimensions may have the same 
spacetime isometry group, theories in different dimensions have the 
potential to be dual. Since the conformal group of $d$-dimensional 
Minkowski spacetime and the isometry group of $(d+1)$-dimensional 
anti-deSitter spacetime are both $SO(d,2)$, there is the potential of a 
large number of dualities between conformal theories on Minkowski spacetime 
and theories on AdS, or asymptotically AdS (AAdS) spacetimes. This of course 
is a first requirement, and it is necessary to establish a correspondence 
between operators, and between states, in the two theories. For example 
collective states and composite operators of one theory may correspond to 
more ``basic'' states and operators in the other. If the duality is to 
aid in performing computations in one theory which are difficult in the 
other, it is necessary to establish the latter correspondences.  

The first example of an AdS/CFT duality was discovered by Maldacena.\cite{malda} 
It is between supergravity on  AAdS spacetimes and 
supersymmetric conformal Yang-Mills theory. This has been extensively 
studied in the last few years\cite{review}. It has the potential to yield 
insights into quantum gravity, since in principle the dual YM theory 
provides a window. In practice many key questions remain unanswered, 
among them the role that spacetime diffeomorphisms play in the YM theory. 
  
One of the results of these works, of interest for this paper, is the 
proposal for determining the expectation value of the Wilson loop in the 
Yang-Mills theory via a computation in supergravity.\cite{mloop} This 
proposal has been studied in detail by a number of authors, \cite{wloop,otherloops} 
and includes discussion of the exciting possibility of correspondences between 
(quantum) phases of Yang-Mills theory and classical AAdS geometries. 

The proposal is the following: consider string worldsheets $s_\gamma$ in 
an AAdS spacetime which have the loop $\gamma$ as boundary.  
The expectation value of the Wilson loop is given by 
\be 
 <W_\gamma> = \int Ds_\gamma\ e^{-S_{NG}( s_\gamma)} 
\ee
where $S_{NG}=\int d\sigma d\tau \sqrt{h}$ is the Nambu-Goto action of the world 
sheet.\cite{mloop,wloop}
(This is reminiscent of the ``no-boundary'' proposal for quantum gravity.)
In practice, the integral is approximated by computing the Nambu-Goto 
action for a solution of the classical string equations, for a choosen 
class of surface $s$, bounding loop $\gamma$, and AAdS spacetime.  

For static rectangular loops, of space and time extensions $L$ and $T$ lying 
along the $t,x$ coordinates of the Minkowski boundary, $<W_\gamma>$ determines 
the quark potential $V(L)$. The calculation 
in fact gives a divergent $S_{NG}(s_\gamma)$. This is renormalised by 
subtracting the divergent part of the integral, which is proportional to the 
loop's circumference. Thus 
\be 
<W_\gamma> = e^{-S_{NG}(L) - k L} \sim e^{-TV(L)},
\ee  
where $k$ is a constant.  

In principle this computation should be capable of yielding any of the three 
phases of YM theory: $V(L) \sim -1/L$ (Coulomb), $V(L) \sim L$ (confinement) 
and $V(L) \sim -e^{-L}/L$ (Higgs) depending on the geometry used. In this way 
the proposal provides a link between AAdS geometries and phases of 
Yang-Mills theory. In particular, it may also provide a means of seeing how horizon 
and singularity information is encoded, or at least interpreted in the conformal 
YM theory. 

The calculation was first done for the global AdS geometry,\cite{mloop} and 
subsequently for the AdS-Schwarzschild metric.\cite{temploop} In the former case, 
one obtains the result expected on grounds of unbroken conformal invariance -- the 
only scale is the quark separation $L$, which results in the Coulomb phase. In the 
latter case, the black hole horizon provides a scale. The calculation for this geometry 
shows a distortion of the Coulomb behaviour such that the the potential 
$V(L)$ goes to zero at a finite value of $L$. This is physically interpreted 
to be the result of temperature, which screens the $-1/L$ behaviour of the potential 
for $L$ larger than a critical ``screening length'' $L_c$. The derivation of this 
result follows.  

The specific form of the AdS black hole metric used is 
\be 
ds^2  = \left({r\over l}\right)^2\left[-f(r)dt^2 + (dx^i)^2 \right] + 
          \left({l\over r}\right)^2 f^{-1}(r) dr^2
\ee 
where $ f(r) = 1 - r_0^4/r^4$, and $r_0$ is the black hole horizon 
parameter.\footnote{ The $S^5$ part of the $10-$dimensional metric and the string 
scale $\alpha'$ also appear in the metric but are not relevant for the main outline 
of the calculation. The AdS scale $l$ is related to the YM parameters $(g,N)$ 
by $l^2 = \sqrt{4\pi gN}$.}  The conformal Minkowski boundary coordinates where 
the loop lies are $t,x^i$ ($i=1..3$). A static 
rectangular loop may be taken to lie along one of the spatial coordinates  
$x$, and $t$. The action is calculated  by setting $\tau = t$ and $\sigma = x$, 
such that $r=r(x)$ describes the embedding of the worldsheet in the black hole 
geometry. The Euclidean metric is used, which gives 
\be 
S_{NG} = T\int dx \sqrt{(\p_x r)^2  + (r^4 - r_0^4)/l^4}
\ee
This is extremised by noting that it doesn't depend explicitly on $x$,  
so the ``energy'' 
\be 
 e = {r^4 - r_0^4 \over \sqrt{(\p_x r)^2  + (r^4 - r_0^4)/l^4} } 
\ee 
is conserved. It is useful to write the constant $e$ as a function of the 
minimum value $r_m$ of $r(x)$. Thus $r_m$ is given by 
$ e = l^2\sqrt{r_m^4 - r^4_0}$.  
Integrating the last equation gives an integral for $x(r)$:
\be 
x(r) = {l^2\over r_m} \sqrt{1-\left({r_0\over r_m}\right)^4} \int_1^{r/r_m}
{dy\over \sqrt{(y^4-1)(y^4 - (r_0/r_m)^4)} }.
\ee 
As $r\rightarrow \infty$ (ie. to the boundary), this integral gives  
half the spatial length $L$ of the loop (since it is symmetric in $r$). 
Thus
\be 
L(r_m) =  {2l^2\over r_m}\sqrt{1-\left({r_0\over r_m}\right)^4} \int_1^{\infty}
{dy\over \sqrt{(y^4-1)(y^4 - (r_0/r_m)^4)} },
\label{L}
\ee 
which relates the loop dimension $L$ to the minimum value $r_m$ of $r$. 
The  action for this solution is 
\be 
S_{NG}|_{{\rm loop}} 
= Tr_m\int_1^\infty dy \sqrt{ {y^4 - (r_0/r_m)^4\over y^4 - 1} }. 
\ee
This integral is divergent and proportional to $y$ for large $y$. It is rendered finite 
in the usual way by integrating up to $y_{max}$, subtracting  the divergence $r_my_{max}$ 
(up to an additive constant to be determined), and then removing the regulator 
$y_{max}\rightarrow \infty$. There is a physical interpretation of this procedure which 
provides the additive constant: 
the subtracted term corresponds to the energy of free quarks, which is also divergent, 
and given by the static configuration where $r$ is not a function of $x$. This solution 
is obtained by setting $\tau = t$ and $\sigma = r$ (with all other coordinates constant) 
in the Nambu-Goto action: $ S_{NG}|_{{\rm free}} = \int_{r_0}^{\infty} dr$,
where the lower limit is the horizon radius, since this is the lowest value of $r$ 
in the Euclidean calculation. The finite potential $V(r_m)$ is therefore
\be 
V(r_m) = r_0 - r_m  
+ r_m\int_1^\infty dy\left(\sqrt{ {y^4 - (r_0/r_m)^4\over y^4 - 1}} - 1\right)  
\label{VL}
\ee
The function $V(L)$ is obtained by substituting $L(r_m)$ from Eq.~(\ref{L}) into the 
last equation, which may be done numerically. Figure 1. shows the typical behaviour: 
the screening length $L_c\sim 0.76$ corresponds to the intersection of the graph with the 
$L$ axis. The small $L$ behavior is $\sim -1/L$, and the result is considered 
unphysical above the $L$ axis. For global AdS  ($r_0 = 0$), $V(L)\sim -1/L$ for all 
$L$. 
\begin{figure}[htbp] 
\vspace*{13pt}
\centerline{\psfig{file=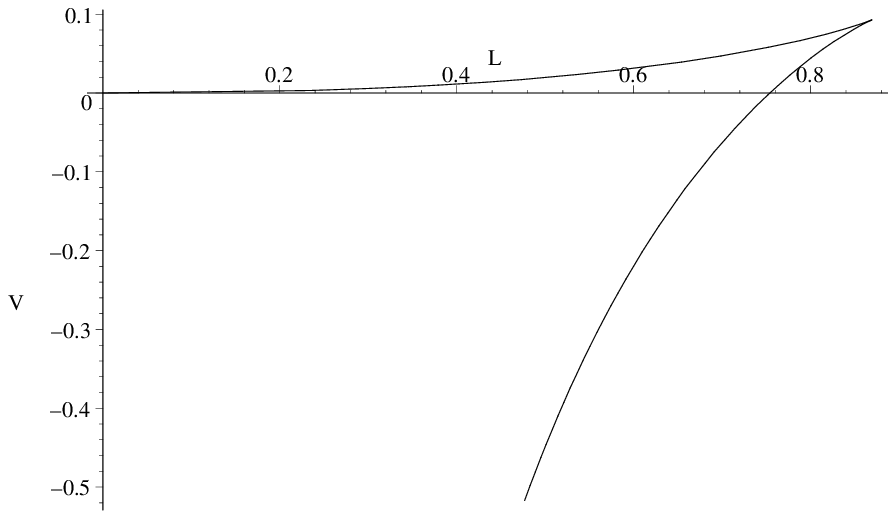}} 
\vspace*{13pt}
\fcaption{$V(L)$ for the positive mass AdS black hole for $r_0=l=1.0$.}
\end{figure}  
 
\section{Naked singularities}
\noindent

From the AdS black hole result, it appears that dimensional parameters 
in metrics will lead to distortions of the Coulomb behaviour of $V(L)$. 
The natural question is whether spacetimes other than black holes can 
lead to qualitatively similar behaviour of $V(L)$, and if so, what the YM 
interpretation should be. This is the question we now probe by using AAdS 
spacetimes that have naked singularities. 

We first give a new solution of the Einstein-scalar field 
equations for massless minimally coupled scalar field with negative 
cosmological constant. This solution is used as the AAdS geometry 
for the Wilson loop calculation, and is compared with the result for 
the negative mass AdS-Schwarzschild metric.

The low energy actions derived from string theory are of the form
\be
S = \int d^dx\ \sqrt{-g} \left[e^{-\phi}\left(R + (\nabla\phi)^2\right) 
+ \Lambda + L_{\rm matter} \right]. 
\ee
Such actions may be put into the standard minimal coupling form by 
the conformal transformation of the metric $g \rightarrow g$exp$(-\phi)$. 
Thus we can consider the usual Einstein-scalar equation in $d$ dimensions
\be 
G_{ab} - {(d-1)(d-2)\over 2l^2} g_{ab} = \p_a\phi\p_b\phi-{1\over 2} 
g_{ab}\p_c\phi\p^c\phi.
\ee
The new solution is obtained starting with the metric ansatz
\be 
ds^2 = \left({r\over l}\right)^2 
\left( -f(r)dv^2 + \sum_{i=1}^{d-2} (dx^i)^2 \right) 
+ 2g(r)dvdr
\ee
where $r,v$ are radial and advanced time coordinates. The solution is 
\be 
f(r) =1, \ \ \ \ g(r) = 
\left[ 1+ \left({r_0\over r}\right)^{2(d-1)} \right]^{-1/2}
\ee
with scalar field given by 
\be 
\p_r\phi(r) \sim  {1\over r}
\left[1 + \left({r\over r_0}\right)^{2(d-1)} \right]^{-1/2}, 
\ee
where the proportionality factor depends on $d$. For $r_0=0$ the solution 
is global AdS in these coordinates -- the scalar field vanishes with $r_0$. 
In all dimensions $d\ge 3$, the large $r$ behavior indicates that the 
spacetimes are asymptotically AdS, with fairly rapid falloff with $r$. 
It is also evident that there is a timelike curvature singularity at $r=0$. 
The conformal Minkowski boundary is obtained by multiplying the solution by 
the factor $(l/r)^2$ and taking the limit $r\rightarrow \infty$. 

We consider the case $d=5$, and describe the Wilson loop calculation using 
world sheets in this geometry. The key similarities and differences from 
the black hole case will become evident. As for the AdS black hole, consider 
the surface given by $r=r(x)$ with $\sigma = x$ and $\tau = v$.  The action is 
\be 
S_{NG} = T \int dx\sqrt{\left( r\over l\right)^4 
+ (\p_xr)^2 \left[1 + \left( {r_0\over r}\right)^8 \right]^{-1} }. 
\ee
On the solution this is 
\be 
S_{NG}|_{{\rm loop}} =T r_m\int_1^\infty {y^6 dy\over \sqrt{(y^4-1)(y^8 + (r_0/r_m)^8)}},
\ee
where $r_m$ is the minimum value of $r(x)$ for this worldsheet. 
The dimension $L$ of the rectangular loop is related to $r_m$ by 
\be 
L(r_m) = {2l^2\over r_m}\int_1^\infty {y^2 dy\over \sqrt{(y^4-1)(y^8 + (r_0/r_m)^8)}}. 
\label{Lsing}
\ee 
\begin{figure}[htbp] 
\vspace*{13pt}
\centerline{\psfig{file=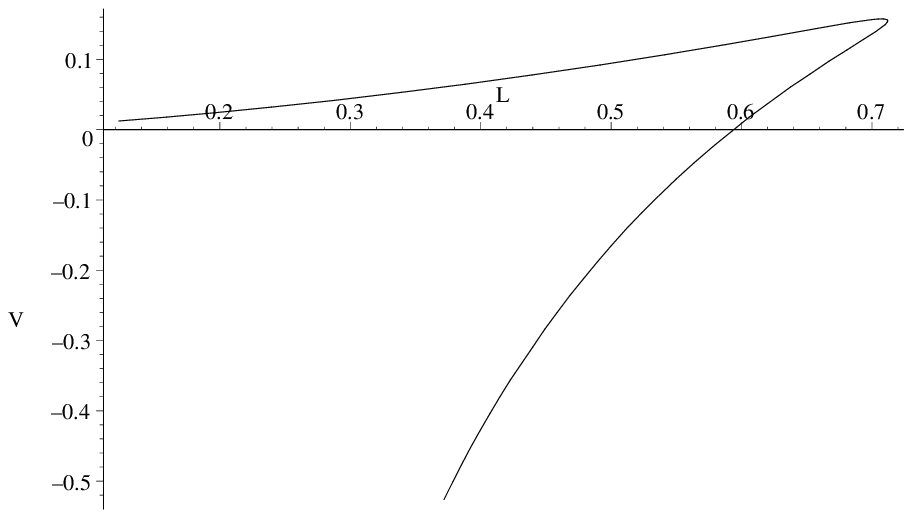}} 
\vspace*{13pt}
\fcaption{$V(L)$ for the singular scalar field solution for $r_0=l=1.0$.}
\end{figure}  
The action and hence $V(r_m)$ are divergent on the solution, with the divergence 
again proportional to $y$ for large $y$. Following the same renormalization 
procedure, we subtract the divergent action corresponding to free quarks. This is 
the configuration  $\tau =v$ and  $\sigma = r$  ($\ne r(x)$) with all the 
$x^i$ constant. Its action is 
\be 
S_{NG}|_{{\rm free}} = T\int_0^\infty dr\ g_{vr} 
= \int_0^\infty dr \left[1+ \left( {r_0\over r}\right)^8 \right]^{-1/2}.
\ee
This differs from the constant integrand obtained for the AdS black hole. The finite 
potential is obtained as before by integrating the loop and free actions up to a 
finite $y_{max}$, subtracting, and taking the limit $y_{max}\rightarrow\infty$. 
This gives 
\be
V(r_m) = r_m\int_1^\infty {y^4 dy\over \sqrt{y^8 + (r_0/r_m)^8}}
\left[{y^2\over \sqrt{y^4-1}} - 1 \right] 
 -r_m\int_0^1dy {y^4\over \sqrt{y^8 + (r_0/r_m)^8}}.
\ee
The last equation combined with Eq. ~(\ref{Lsing}) gives $V(L)$ for 
the scalar field solution. A typical plot of $V(L)$ is shown in Fig. 2. 
The similarity with the AdS black hole case is apparent, in particular  
the intersection with the $L$ axis at a critical $L_c$. Also similar 
is that $V(L)$ becomes double valued above this axis, which again 
indicates that the calculation is not valid for $L>L_c$.  

The comparison of these results with the negative mass 
AdS-Schwarzschild case is interesting.\footnote{I thank T. Padmanabhan 
for asking about this case.} The calculation proceeds in exactly 
the same way with the only difference being the subtraction integral necessary 
to obtain a finite $V(L)$: this now has lower limit $r=0$ rather than the 
horizon radius $r=r_0$. The result is 
\bea 
V(r_m) &=& r_m\left[ -1 
+ \int_1^\infty dy\left(\sqrt{ {y^4 + (r_0/r_m)^4\over y^4 - 1}} - 1\right)\right],\\
L(r_m) &=&  {2l^2\over r_m}\sqrt{1+\left({r_0\over r_m}\right)^4} \int_1^{\infty}
{dy\over \sqrt{(y^4-1)(y^4 + (r_0/r_m)^4)} }.
\eea
The graph of $V(L)$ appears in Fig. 3. The features similar to the previous 
cases are again the $-1/L$ behaviour for sufficiently small $L$, and the 
intersection with the $L$ axis. The additional feature is that the potential 
does not become double valued for $L>L_c$, and thus may represent real physics. 
Remarkable is the linear behavior in this region, which indicates a confining 
potential. Thus this singular spacetime seems to correspond to the Coulomb to 
confining phase transition!
  
 \begin{figure}[htbp] 
\vspace*{12pt}
\centerline{\psfig{file=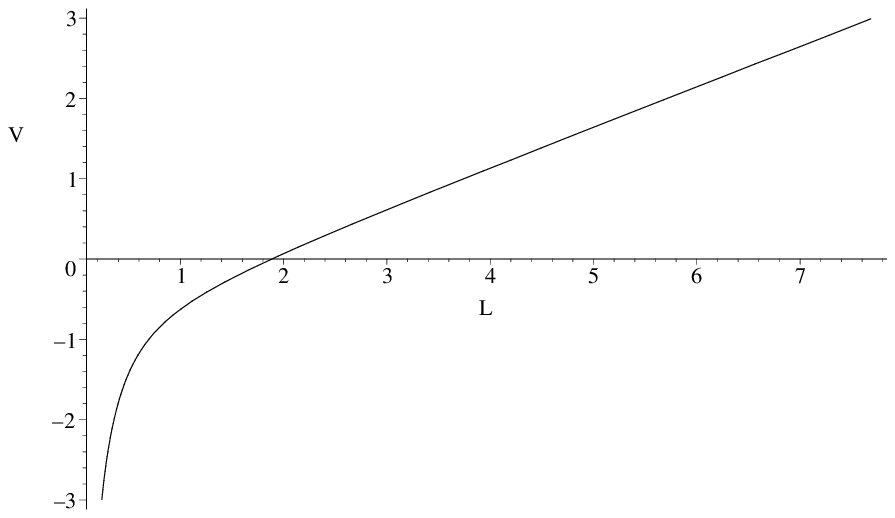}} 
\vspace*{13pt}
\fcaption{$V(L)$ for the negative mass AdS black hole for $r_0=l=1.0$.}
\end{figure}  

\section{Conclusions}
\noindent
We have seen that there are explicit dualities between a class of topological 
field theories and the simple harmonic oscillator. These appear to be mainly of 
mathematical interest since the correspondences are in a sense too limited to 
extract ``physics.'' The model nevertheless raises the question of what physics 
can be learned about geometrical theories from their non-geometrical  
quantum duals. 

The purpose of the AdS/CFT Wilson loop calculation for singular spacetimes 
is to see how their metric parameters are manifested in the YM quark potential. 
A component of this question is whether the calculation even leads to sensible 
results. 

The surprising answer is that the distortion of the Coulomb behaviour  
for the AdS black hole is also present in the singular scalar field 
and negative mass AdS-Schwarzschild spacetimes, and it is qualitatively very 
similar: a ``thermal screening'' interpretation is clearly possible by looking 
at just the potential for $L\le L_c$. 

The negative mass AdS-Schwarzschild metric presents an additional surprise: 
it suggests a Coulomb to confining phase transition at $L=L_c$. Why a nice 
result like this follows from such a bad spacetime requires further probing 
of the AdS/CFT conjecture.   

\noindent{\it Note added:} After this work was submitted to the archives,
additional references \cite{sabra,greg,alvarez} concerning Wilson loop 
calculations were brought to the author's attention. The first paper describes 
a supersymmetric AAdS solution which could be used to study the Wilson loop.
The second and third papers give Wilson loop calculations respectively 
for the charged AdS black hole, and a certain regular bosonic background. 
The charged black hole metric also gives a screening potential, suggesting 
that the addition of charge does not alter the results, wheras the latter 
metric gives a linear potential.

\nonumsection{Acknowledgments}
\noindent
I thank the organisers for their invitation to speak at the Workshop,  
and the participants for their questions and comments. I also 
thank Saurya Das and Lee Smolin for their comments on the manuscript. 
This work was supported in part by the Natural Science and Engineering 
Research Council of Canada.

\nonumsection{References}

\end{document}